%% file: main.tex
\documentclass[runningheads]{llncs}
\usepackage{graphicx}
\usepackage{multirow}

\usepackage{stmaryrd} % crochet entier pour math
\usepackage{url} % url
\usepackage{booktabs} % bottomrule for table
\usepackage[inline]{enumitem}
\begin{document}
\title{A Study on FGSM Adversarial Training for Neural Retrieval}
\author{Simon Lupart \and
Stéphane Clinchant}
%Third Author\inst{3}\orcidID{2222--3333-4444-5555}}
%
\authorrunning{S. Lupart and S. Clinchant}
\institute{Naver Labs Europe, Meylan, France % \and
%Springer Heidelberg, Tiergartenstr. 17, 69121 Heidelberg, Germany
\email{\{simon.lupart,stephane.clinchant\}@naverlabs.com}}
\maketitle              % typeset the header of the contribution
\begin{abstract}

Neural retrieval models have acquired significant effectiveness gains over the last few years compared to term-based methods. Nevertheless, those models may be brittle when faced to typos, distribution shifts or vulnerable to malicious attacks. For instance, several recent papers demonstrated that such variations severely impacted models performances, and then tried to train more resilient models. Usual approaches include synonyms replacements or typos injections -- as data-augmentation -- and the use of more robust tokenizers (characterBERT, BPE-dropout). To further complement the literature, we investigate in this paper adversarial training as another possible solution to this robustness issue. Our comparison includes the two main families of BERT-based neural retrievers, i.e. dense and sparse, with and without distillation techniques. We then demonstrate that one of the most simple adversarial training techniques -- the Fast Gradient Sign Method (FGSM) -- can improve first stage rankers robustness and effectiveness. In particular, FGSM increases models performances on both in-domain and out-of-domain distributions, and also on queries with typos, for multiple neural retrievers. 
%Finally, from the observation that FGSM adversarial training could be use as a regularization tool, we test it on several BEIR datasets in domain-adapation scenarios, when data and training ressources may be more scarse.

% Pretrained Language Models (PLM) have recently shown their strength in IR, outperforming their predecessors both on In-Domain (ID) and Out-Of-Domain (OOD) distributions. 
% However, the behaviour of those models under query or document perturbations is still an issue, in particular when typos, naturality variations or paraphrasing occur. In the following paper, we investigate adversarial training as a solution to such scenarios, and compare adversarial training objectives in the case of IR. Our analysis demonstrates that simple adversarial training techniques such as Fast Gradient Sign Method (FGSM) can improve first stage rankers robustness. We also discovered that the adversarial component was beneficial on ID and OOD distributions for multiple state-of-the-art neural rankers such as dense bi-encoder and SPLADE. Finally, we also compare the adversarial component with distillation techniques, to see to which extend they can be combined. Our code is available at githublink.
% The abstract should briefly summarize the contents of the paper in
% 15--250 words.

\end{abstract}

\section{Introduction}

Stochastic Gradient Descent (SGD) is the main optimization method in Machine Learning, enabling to effectively optimize neural networks with millions of parameters. Despite the great performances from SGD, neural networks models still suffer from robustness issues when face to distributions shifts or noise.
% and are brittle when face to distributions shifts. 
The seminal work of Goodfellow et al.~\cite{goodfellow2015explaining} showed, for instance, how to manipulate model-predictions -- in an adversarial way -- by adding gradient-targeted perturbations in images at the pixel level. Their approach, the Fast Gradient Sign Method (FGSM), was the first and simpler algorithm to perform such attack. While this opened the way to possible stronger attacks, it was shown in the meantime, that the same techniques could also be used to train more robust and resilient models. Beyond original attacks, Adversarial Training (AT) could be used to increase model robustness, as a regularization or data-augmentation~\cite{https://doi.org/10.48550/arxiv.2105.12932}.
%As to find weaknesses of models, people have thus intuitivelly searched for approaches that would also use gradients. The Fast Gradient Sign Method (FGSM) is the most known gradient-technique of such. 

%Introduce
While in the field of Information Retrieval (IR), several works demonstrated that Pre-trained Language Models (PLM) based architectures had the same robustness issues in zero-shot and noisy environments~\cite{beir,qvg,Zhuang_2022}, it seems to our knowledge that one of the simple adversarial training technique -- FGSM-AT -- has not been evaluated for first stage rankers. 
% In particular for the dense bi-encoder , it was recently shown that some dense bi-encoder models (DPR, tas-b) had severe performance drop when face to distributions shifts, as shown in the BEIR benchmark.
%Since Pre-trained Language Models (PLM)  are now  the new paradigm in Information Retrieval, we want to apply adversarial training on neural retrieval architectures, as it was recently shown that some dense   models (DPR, tas-b) had robustness issues.
As an initial study, we consider in this work FGSM-AT, both to increase model robustness both for in and out-of-domain. Then, we apply FGSM-AT on domain adaptation scenarios, to further analyse AT in environments with fewer annotated samples.
% With the recent advances in Information Retrieval and the use of Pre-trained Language Model as backbone of neural retrieval, 
% Besides, it we know that with the recent advance of PLM, lots of dense models were shown not to be robust with perf OOD lower that term-based ones (BM25)
% Robustness being an major issue in Information Retrieval, as it was shown that bert-based dense models (for instance DPR, tas-b) had robustness issues
% we know that with the recent advance of PLM, lots of dense models were shown not to be robust with perf OOD lower that term-based ones (BM25). With the specificity of IR being that most of documents and queries come from the web, it makes retriever models especially exposed to noise.
% Our study stands exaclty in this context, using FGSM to improve performances of neural retrievers.
% Due to the direct contact 
% Adversarial attacks in Information Retrieval (IR) is 
% Neural first stage rankers (bi-encoder, colbert, splade) are the new meta in IR, because they are much more effective than BM25 both in domain and out of domain. In the process of first stage ranking (at inference), models creates representations for millions of docs and queries from the web, and thus are prone to adversarial attacks.
%The following of the paper will be organized as to answers to the following Research questions considering FGSM:
Overall, this paper investigates the following Research Questions (RQ):
\begin{itemize}
    \item RQ1: How performances change on in-domain and out-of-domain distributions with FGSM Adversarial Training?
    \item RQ2: Does FGSM Adversarial Training increase performances in environments with noise in queries such as typos?
    \item RQ3: Is FGSM Adversarial Training beneficial for domain adaptation?
    % \item What is the best objective for the adversarial training perturbations? self-teaching? or to considering them as independent samples?
    % \item Is targeted adversarial training on query or document side beneficial in specific scenarios? (cf BEIR; dataset by dataset)
    % \item we discovered that models have difficulties for matching on ood words, can we create an eval dataset of this, and see if we can improve the exact match on new words? +atomic metric (paper we read in Reading Group)
\end{itemize}

% jeu de données principal = beir, celui sur lequel on fait le choix de la loss et du epsilon de la perturb. Puis ablation study (query/doc side) sur qvg et beir, puis general perf sur msmarco/trec. Cela couvre les perturbed queries, le cas ood et in domain

\section{Related Works}

There is an abundant literature on adversarial methods, which can be grouped in mainly two families: the white-box and the black-box methods. In the white-box settings, one assumes full access to the model and can therefore compute models gradients (e.g., FGSM, PGD~\cite{goodfellow2015explaining,https://doi.org/10.48550/arxiv.1706.06083,https://doi.org/10.48550/arxiv.1802.00420}), in difference to the black-box settings, where gradients are hidden from the attacker. %(those techniques are also refered as white box techniques, because you consider that you have access to all information from the model), 
In particular for the black-box case, attacks thus rely on various heuristic techniques, by iterating on the models inputs/outputs.
While white box settings apply well in Computer Vision, examples of black box attacks are more common in NLP due to the discrete nature of words. For instance, BERT-Attack~\cite{bert-attack}, iteratively replaces words by their synonyms -- using a MLM BERT head -- to find possible replacement-words that could trigger the model to make wrong predictions. 
To further specify the literature on adversarial methods, some works purely focus on malignant objectives~\cite{https://doi.org/10.48550/arxiv.2106.09667}, while others try to overcome the weaknesses of current architectures (Adversarial Training). As an example of the former, Carlini et al.~\cite{poison} show that by poisoning a minimal fraction of the training set, we could control the prediction of particular test samples.
% In the latter, the objective is to create more robust models through optimization (e.g. GAN, AR2~\cite{zhang2022adversarial}), without any malignant intent. 
% In this case, white-box scenario is justified because you are training the model, so you have access to the gradient.

With the emergence of PLM-based models in IR (dense bi-encoder, SPLADE, ColBERT~\cite{karpukhin2020dense,spladev2,colbertv2}) replacing old term-based approaches (BM25~\cite{robertson2009probabilistic}), some literature also appeared on adversarial methods in IR.
In the current literature, the first works focus on malignant attacks, also known as Search Engine Optimization (SEO)~\cite{https://doi.org/10.48550/arxiv.2206.11724,https://doi.org/10.48550/arxiv.2204.01321,Wu2022CertifiedRT}. Applied to IR, the goal becomes to either promote or demote the rank of a particular document or set of documents. As a leverage, existing works usually add several tokens in a document, that are optimized to modify its rank for a given query, or a set of queries. Distillation being also very commonly used in IR~\cite{distil_hof}, grey-box approaches also appeared. In their work, \cite{https://doi.org/10.48550/arxiv.2209.06506} present the idea as to first distil a model -- on which we would not have access to the gradient -- into a copy, and then attack through the gradient of the copy.

Although the literature on SEO is already rich, it appears that adversarial training in IR is very limited, to our knowledge. Zhuang et al.~\cite{https://doi.org/10.48550/arxiv.2108.12139} used data-augmentation on typos to make models more robust to typos. Later, the same authors proposed a dedicated architecture for typos~\cite{Zhuang_2022}: their model used CharacterBERT, and a smoothing technique they called Self-Teaching, which forces the model to predict the same score for a given (query, document) pair with/without the typos. In the meantime, Sidiropoulos et al.~\cite{Sidiropoulos_2022} also experimented with data-augmentation and contrastive losses between queries with and without typos, and had similar results. In the following of the paper, we aim at applying the same methods with perturbations directly injected in the embedding space -- in difference to previous works that worked at the token level -- and also with adversarial perturbations.

\section{Adversarial Training}
\label{adv-train}
This section introduces adversarial training for first stage rankers in IR, using the most simple approach, i.e. FGSM-AT.
Standard training in IR usually uses a contrastive \texttt{InfoNCE} loss~\cite{infonce} on triplets $T_i=(q_i, d_i^+, d_i^-)$, which aims at increasing the similarity between the query and the positive document, while reducing it for the negative documents. It can be seen as minimizing the loss: 
$$
\mathcal{L}_{\mathtt{InfoNCE}}(T_i)=- \frac{e^{s(q_i, d_i^+)}}{e^{s(q_i, d_i^+)} + \sum_j e^{s(q_i, d_{i,j}^-)}}
$$
% + \sum_j e^{s(q_i, d_{i,j}^+)}}  + e^{s(q_i, d_i^-)}}
Now in an adversarial training scenario, each triplet is perturbed by an $\epsilon_i=(\epsilon_{i_q}, \epsilon_{i_{d^+}}, \epsilon_{i_{d^-}})$, containing independent perturbations for the query, and each of the documents (applied on the inputs embeddings). Then, to ensure that the model would predict the same scores in a local vicinity around any training triplet, FGSM-AT minimizes the joint objective containing the original and adversarial losses as follows\footnote{Note that FGSM-AT can be applied on any loss, and thus generalizes to the margin-MSE loss for the case of distillation~\cite{distil_hof}.}:
$$
\mathcal{L}_{\mathrm{total}}(T_i) = \mathcal{L}_{\mathtt{InfoNCE}}(T_i) + \mathcal{L}_{\mathrm{adv}}(T_i+\epsilon_i) 
$$
$$
\epsilon_i =\mathrm{argmax}_{||r||_2 \leq ||r_{\mathrm{max}}||_2} \; \mathcal{L}_{\mathrm{adv}}(T_i+r)
$$
where $\mathcal{L}_{\mathrm{adv}}$ is the adversarial loss, either the original $\mathcal{L}_{\mathtt{InfoNCE}}$ ranking loss -- which is the case we consider for the following of the paper -- or a measure of divergence on scores directly (e.g., Kullback Leibler Divergence between the distributions of scores).
% $L_{adv}(T_i+\epsilon_i) = D[s(T_i) || s(T_i+\epsilon_i)]$.
Note that adversarial training can be defined with a norm (here $||.||_2$) and an upper-bound on the norm (here $r_{\mathrm{max}}$).
With FGSM-AT, the min-max optimization process is simplified by approximating $\epsilon_i$ in one step, computing the gradient with respect to the input at $T_i$, and taking the direction that maximizes it. The norm of the perturbation is also constant ($||\epsilon_i||_2=||r_{\mathrm{max}}||_2$):
% The norm is also taken fix ($r=r_{max}=0.01$):
$$
\epsilon_i = - r_{\mathrm{max}} \frac{g_i}{||g_i||_2} \qquad g_i = \nabla_{T_i} \mathcal{L}_{\mathrm{adv}}(T_i).
$$
Intuitively, this helps the model to smooth the representation space, and act as a regularization. From another perspective, this can also be seen as a data augmentation, as we simply create one new sample for each original sample.

\paragraph{\bf{Baselines.}} As baselines for FGSM-AT, we compare it with a \textbf{$\epsilon-$random} baseline that adds a perturbation in a random direction of the embedding space (instead of the one given by the gradient), and a token level baseline \textbf{\texttt{token}-random}, that replace 15\% of the original tokens with other random tokens. 
Finally, we also report performance from an \textbf{universal} adversarial training baseline~\cite{univ0,univ1,univ2}. In difference to FGSM-AT that uses a distinct perturbation for each triplets, universal-AT considers the same embedding perturbation $\mathbf{\epsilon_{all}}$ for all triplets. The perturbation can be thus optimized at the global level, as an additional parameter of the model with the opposite objective (i.e. maximizing the loss). While this baseline is usually weaker than FGSM-AT, it is more efficient than its counterpart because it requires only one back-propagation per triplets, instead of the two required by FGSM-AT.

\section{Experiments}

%\subsection{Experimental Settings}

We compare two neural retrieval architectures: \begin{enumerate*}[label=(\textit{\roman*})]
    \item a \textbf{dense bi-encoder}, which uses dot products to compute similarities between the \texttt{mean} tokens representations of queries and documents \cite{karpukhin2020dense},
    \item and \textbf{SPLADE} -- as a sparse bi-encoder -- which represents them as high-dimensional bag-of-words vectors \cite{splade++formal}.
    % \item and ColBERT, which uses the token-by-token representations in a late-interaction manner.
\end{enumerate*}
Both models are trained on MS MARCO. The dense bi-encoder is trained for 5 epochs, over the full set of 500k queries, in batches of size 16 with 32 negatives per query (with the hard-negatives released by \cite{splade++formal} from SPLADE-Cocondenser). For SPLADE, we use the same process, but with batches of sizes 8 and harder negatives.
% Models are trained on MS MARCO triplets for 150k iterations. For all FGSM experiments, we additionally train for 60k iterations with the targeted perturbation (following the settings from \cite{noise-inject}). While our dense bi-encoder -- with and without distillation -- is trained using a batch size of 4 with 32 negatives for each positive, SPLADE is trained with only a single negative per query. SPLADE batch size is set to 128. 
For distillation, the dense bi-encoder uses a released msmarco-hard-negatives dataset\footnote{https://huggingface.co/datasets/sentence-transformers/msmarco-hard-negatives} hosted on the Transformers library~\cite{sentencebert} where negatives were scored by a larger reranker, while SPLADE uses the negatives than previously, scored by an ensemble of rerankers. We kept the same batch sizes and numbers of negatives during distillation..
% As previously, the dense was trained with 16-sized batches, where each query had 32 negatives, while SPLADE was trained with larger batches, but with only one negative.
Both models are trained with In-Batch-Negatives and a learning rate of 2e-5 with linear scheduler.
To add FGSM-AT, we start from the previous best checkpoint, and resume training for 2 epochs with the targeted perturbations. This follows the settings from \cite{noise-inject}, with the motivation that FGSM-AT or noise injection can help to recover from a sharp minimum. Also, including FGSM-AT for only the last steps reduces training cost in comparison to FGSM-AT from scratch (each step of FGSM-AT being twice longer than a regular step). The value of the perturbation norm $r_{\mathrm{max}}$ is fixed to $0.01$ (best value from $0.1$, $0.01$, $0.001$ in our initial study). 

\paragraph{\bf{Datasets.}} The models are then benchmarked on the MS MARCO collection~\cite{msmarco}, with both the original MS MARCO dev queries and TREC DL 2019/2020 judgements. To evaluate on out-of-distribution condition, we use the 13 avalaible datasets from the BEIR benchmark~\cite{beir}. Metrics are the default ones, MRR@10 and Recall@1k for MS MARCO, and nDCG@10 for TREC and BEIR datasets.

\subsection{RQ1: How performances change on in-domain and out-of-do- main distributions with FGSM Adversarial Training?}
\input{id.tex}
\label{sec:rq1}
Table~\ref{tab:all} reports the general comparison of FGSM-AT in-domain (MS MARCO and TREC), together with the mean nDCG@10 score out-of-domain on BEIR. The first eight rows report the performances on both the dense bi-encoder and SPLADE without distillation, while for lower rows, the comparison is made on models trained with distillation. Without distillation, the two random baselines first reveals that FGSM-AT is more effective than random noise injection (both at the token level or in the embeddings).
%and that the gradient direction helps finding irregularity in the embedding space.
We then notice the high improvements from FGSM for the dense model (+2.25 on MS MARCO), but also for the sparse model (+1.81) in-domain. As expected, universal-AT places itself in the middle of the two, as a trade-off between effectiveness and efficiency. For models with distillation, we observe that there is an improvement on MS MARCO dev MRR@10, in particular for SPLADE-D, but this improvement is more contested for the dense bi-encoder. For fair comparison, we mention here that we kept similar FLOPS (with and without FGSM) in the case of SPLADE. Now looking at performances out-of-domain, we have a high increase on models without distillation with FGSM on the 13 BEIR-datasets. This gain seems to saturate for the distilled models. Overall, both FGSM and Universal-AT have a very similar behaviour on the dense and sparse architectures, as a proof of their consistency. 

From the observations made on the distilled models, we hypothesis that FGSM-AT and distillation have both a similar label smoothing effect: through the distilled scores and the MSE-loss for distillation, and through the adversarial perturbations for FGSM. This would explain the mixed gains in this case, and why performance increases do not add up. However, note that FGSM-AT smooths representations without requiring external knowledge from a reranker, in difference to distillation.

% Another interesting results is that SPLADE+FGSM is getting close to SPLADE\_distil, without the need of any external information (from the distilled reranker). 

% add some baseline = random token replacement? see how to include them for the qvg dataset - no more space in the table. 

% add a paragraph about how it doesn't work with distillation...

% section about training efficiency

% for arxiv version: include universal training as answer to efficiency + some thoughts on working on the embedding space instead than at the token level with citation \cite{embVStoken}

\subsection{RQ2: Does FGSM Adversarial Training increase performances in environments with noise in queries such as typos?}

\input{typos.tex}

%Our first Research Question gives an overview of FGSM in a clean environment, however, we don't know what would happen in more realistic environment, thus the need for our second RQ.
For the second research question, we examine the effect of adversarial training on queries with typos. To do so, we evaluate our models on the queries variations dataset~\cite{qvg}, based on TREC DL 19.
Table~\ref{tab:typos} contains variations in queries that do not change the semantic of the original query, but apply noise on it, with \textbf{typos} (rows \texttt{b/c/d}), \textbf{paraphrasing} (\texttt{h/i/j/k}) and changes in the \textbf{word ordering} (\texttt{g}) or the \textbf{naturality} (\texttt{e/f}). First, independently from FGSM-AT, we can observe the important drops in all categories, especially for the typos. On typos, SPLADE seems to be naturally more robust than the dense (+8.84 in average without distillation), even-though drops are really important for both models. The better performances of SPLADE may be due to the natural robustness brought by the MLM head.

Now on FGSM-AT, our observation is that, while FGSM-AT is a general method (not a priori focus on one type of noise), it helps in almost all cases. In particular we see gains on paraphrasing for all models, and even on queries with typos for the dense. Due to the small number of queries, lots of $p$-values are over 0.05, however, by computing the mean per category, we observe -- while not reported in the table -- that D-N, S-N and D-D have statistically significant increases for paraphrasing ($p$-values $< 0.05$). This suggest that representations of models trained with FGSM-AT are more robust, and queries with the same intent will be closer to each other.

\subsection{RQ3: Is FGSM Adversarial Training beneficial for domain ada- ptation?}
\input{beir.tex}
As a final research question, we consider the case of scarce training data, through the example of domain adaptation, to investigate if FGSM-AT could mitigate overfitting of pre-trained IR models. For this experiment, we start from the previous distilled dense bi-encoder (D-D) trained on MS MARCO, and finetune it with negative training triplets from resp. datasets from the BEIR benchmark (as for the experiments in \ref{sec:rq1}, we sampled 32 negatives per query from SPLADE-Cocondenser, and also used lower learning rates for adaptation). Only few of BEIR datasets have actual train/dev/test sets which is why we perform our experiment on Fever, FiQA and NFCorpus (containing resp. 110k, 5.5k and 2.6k training queries). Training is done in 100 epochs for FiQA and NFCorpus, and 10 epochs for Fever, with the best checkpoint being selected using the dev set. Training sets being relatively small, we need to train models with a high number of epochs, which is our motivation for using FGSM-AT on this particular settings to smooth representations. Another motivation is that training a reranker for distillation is challenging with only few training samples, and also distillation would require to retrain a reranker for each of the new domain, which is expensive.

Table~\ref{tab:beir_adapt} reports the finetuning results. First, we notice that the distilled bi-encoder -- initially trained on MS MARCO with distillation -- is able to learn from the new BEIR annotations, in particular on Fever and NFCorpus (+7.48 and +8.88 nDCG@10 resp.), and overall that FGSM-AT prevents the models from overfitting. Results of FGSM-AT are different on FiQA, but this dataset is also the one on which models have the most struggle to learn from the new annotations (gains from only +2.28), so the different behaviour may be due to poor training data, more than FGSM-AT in itself.

\section{Conclusion}
In this study, we experimented with FGSM to train first stage rankers. Our experiments revealed that a simple regularization on the embedding space could increase the in-domain performances  on MS MARCO, especially for models trained without distillation, on which it additionally strengthen the generalization capacities. Besides, FGSM-AT enables a better adaptation to new domains, even on top of distilled models. In future work, we plan to investigate adversarial training directly on rerankers to see if improvements on rerankers could transfer during distillation. Finally, we hope our study would encourage the community to reconsider this baseline method when dealing with robustness issues.

\bibliographystyle{splncs04}
\bibliography{mybib}

\end{document}

%% file: id.tex
\begin{table}[t!]
\centering
    \caption{In-domain performances on MS MARCO dev et TREC DL tracks, and out-of-domain average performance on the 13 BEIR datasets. We report scores for both standard negative training (-N) and distillation training (-D). Results with $\dag$ indicates $p$-values $< 0.05$ on paired t-test.}
    \label{tab:all}
    \begin{tabular}{l  c c c c c c | c}
        \toprule
        \textbf{Dataset ($\rightarrow$)} & \multicolumn{2}{c}{MS MARCO dev} & \multicolumn{2}{c}{TREC DL 2019} & \multicolumn{2}{c|}{TREC DL 2020} & BEIR(13) \\
        \textbf{Models ($\downarrow$)} & MRR@10 & R@1k & nDCG@10 & R@1k & nDCG@10 & R@1k & nDCG@10 \\ \midrule 
        \midrule
        % \multicolumn{2}{c}{Negative Training} \\ \midrule
        bi-encoder -N & 33.24 & 95.75 & 65.93 & \textbf{76.05} & 66.08 & 81.11 & 39.54 \\ 
        +\texttt{token}-random &33.21	&95.42	&65.61	&75.12	&66.26	&79.76	&39.09 \\
        % +\texttt{token}-MLM &33.34	&95.38	&65.85	&75.18	&64.46	&80.7	&38.78 \\
        +$\epsilon-$random & 32.98 & 95.52 & 64.99 & 75.51 & 65.34& 79.66 & 38.68 \\
        +Universal & $34.17^\dag$	&96.03	&66.32	&76.77	&67.71	&80.41	&40.36 \\
        +FGSM &  $\textbf{35.49}^\dag$	 & $\textbf{96.38}^\dag$	 & $\textbf{69.24}^\dag$	 & 75.57 & $\textbf{68.89}^\dag$	 & \textbf{81.32} & \textbf{41.63} \\ \midrule
        % SPLADE -N & 34.59	&96.50&	68.56&	79.65&	67.55&	83.60& 43.97 \\ 
        % +FGSM  & $\textbf{36.02}^\dag$	&$\textbf{96.97}^\dag$&	\textbf{70.13}&	$\textbf{82.44}^\dag$ &	\textbf{69.18}&	$\textbf{86.61}^\dag$ &\textbf{45.51}\\ 
        SPLADE -N & 36.11	&97.35	&71.04	&83.00	&70.22	&85.32& 46.86 \\ 
        +Universal & $36.87^\dag$&97.42	&71.65	&84.22	&\textbf{71.92}	&86.19 & 46.99 \\
        +FGSM  & $\textbf{37.92}^\dag$	&\textbf{97.79}	&\textbf{72.94}	&\textbf{85.27}	&71.86	&\textbf{86.79} &\textbf{47.94}\\ \midrule
        % ColBERT & 36.71&	96.75 &-&-&-&-&-\\ 
        % +Adv & \textbf{36.78}&	\textbf{97.12} &-&-&-&-&-\\ \midrule
        \midrule
         % \multicolumn{2}{c}{Distillation Training} \\ \midrule
        bi-encoder -D & 37.13 & \textbf{97.43} & 71.08 & \textbf{81.18} & 69.68 & \textbf{83.95} &45.13\\  %$\textbf{97.43}^\dag$ $\textbf{83.95}^\dag$
        +Universal &37.34	&97.54	&72.54	&82.43	&70.71	&84.33	& \textbf{45.20} \\
        +FGSM & $\textbf{37.49}^\dag$ & 97.20 &  \textbf{71.42} &  79.80 & \textbf{70.32} & 82.97 & 44.90 \\ \midrule
        % SPLADE -D & 37.05&	\textbf{97.89}&	72.99&	85.62&	70.05&	88.77& \textbf{49.55}\\ 
        % +FGSM  & $\textbf{37.51}^\dag$ &	97.78&	\textbf{73.49}&	\textbf{86.40}&	\textbf{70.89}&	\textbf{88.84} & 49.41\\
        SPLADE -D &40.00	&98.14	&\textbf{76.21}	&88.25	&73.37	&89.44 & 47.52\\
        +Universal &40.08	&\textbf{98.29}&	75.89&	\textbf{88.59}&	\textbf{73.84}&	89.35 &\textbf{47.59}\\
        +FGSM &$\textbf{40.55}^\dag$	&98.22	&76.13&	87.84&	73.66&	\textbf{89.55}&47.44 \\
        \bottomrule
    \end{tabular}   
\end{table}

% \begin{table*}[t!]
%     \caption{In-Domain performances on MS MARCO dev et TREC DL tracks, and Out-of-Domain average performances on the 13 BEIR datasets. We report scores for both standard and distillation models. +Adv corresponds to the models trained with FGSM.}
%     \label{tab:all}
%     \begin{tabular}{l | c c | c | c | c}
%         \toprule
%         Dataset ($\rightarrow$) & \multicolumn{2}{c|}{MS MARCO dev} & TREC DL 2019 & TREC DL 2020 & BEIR \\
%         Models ($\downarrow$) & MRR@10 & Recall@1k & nDCG@10 & nDCG@10 & nDCG@10 \\ \midrule \midrule
%         SIAM & 3116	& 9484&	6415&		6525&	38644 \\ 
%         +Adv &  3230&	9469&	6456&	6391&39546\\ \midrule
%         SPLADE & 3459	&965&	6856&	6755&	 439722\\ 
%         +Adv  & $3602^\dag$	&$9697^\dag$&	\textbf{7013} &	\textbf{6918}&	\textbf{455124}\\ \midrule
%         ColBERT & 3671&	9675 &-&-&-\\ 
%         +Adv & \textbf{3678}&	\textbf{9712} &-&-&-\\ \midrule
%         \midrule
%         SIAM\_distil & 3268&	9575&	6800& 6651 &40828\\ 
%         +Adv & $3333^\dag$&	$9551^\dag$&	6802&	6468&	40784 \\ \midrule
%         SPLADE\_distil & 3705&	\textbf{9789}&	7299&	7005& \textbf{495509}\\ 
%         +Adv  & $\textbf{3751}^\dag$ &	9778&	\textbf{7349}&		\textbf{7089}&494117\\ \midrule
%         \bottomrule
%     \end{tabular}   
% \end{table*}

%% file: typos.tex
\begin{table}[t]
    \caption{Robustness to variation from the query-variation generator dataset - on TREC DL 2019 queries (nDCG@10). D-N and D-D are resp. the non-distil/distil version of dense bi-encoder. The same notation is used for SPLADE with S-N and S-D.}
    \label{tab:typos}
    \centering
    \begin{tabular}{c | l  c c  c c | c c  c c}
        \toprule
        \# & \textbf{Q-Variation} & \textbf{D-N} & \textbf{+FGSM} & \textbf{S-N} & \textbf{+FGSM} & \textbf{D-D} & \textbf{+FGSM} & \textbf{S-D} & \textbf{+FGSM}\\ \midrule
        % \midrule #71.12 76.42 72.88 76.48
        a & Original& 65.93&$\textbf{69.24}^\dag$	 &71.04&\textbf{72.94}	&71.08 & \textbf{71.42}&\textbf{76.21}	&76.13 \\ \midrule
        b & RandomChar&38.82	&\textbf{41.93}	 & 44.40	&\textbf{46.39}	&45.52 & \textbf{47.33}&46.45	&\textbf{46.90} \\ % $\textbf{49.66}^\dag$
        c & NeighbChar& 36.35&	$\textbf{41.18}^\dag$	 & 46.57&\textbf{49.62}	&48.86 &\textbf{49.11}&50.93	&\textbf{52.03}\\
        d & QWERTYChar& 34.43	&$\textbf{40.87}^\dag$	& 45.15&\textbf{46.20}&46.29 & \textbf{48.01}&49.34	&\textbf{50.10} \\ \midrule
        e & RMStopWords& 63.2	&$\textbf{66.90}^\dag$	& 71.45	&\textbf{72.67}	&70.46 & \textbf{70.90} &73.26	&\textbf{74.39} \\
        f & T5DescToTitle& 59.34	&$\textbf{63.10}^\dag$	 & 63.35	&\textbf{66.10}&64.28 & \textbf{65.92}&65.29&\textbf{66.21} \\ \midrule
        g & RandomOrder& 65.81	&\textbf{67.62}	 & 69.69&\textbf{72.28}	&70.76 & \textbf{71.36}&\textbf{75.06}&74.89 \\ \midrule
        h & BackTransla& 58.06	&\textbf{61.29} & 57.84	&\textbf{60.40}	&61.36 & \textbf{63.78}&\textbf{63.78}	&63.29 \\
        i & T5QQP & 63.84&	\textbf{64.62}& 66.46&	\textbf{69.51}&	\textbf{69.50} & 68.36&\textbf{72.09}&	71.63\\
        j & WordEmbSyn& 60.30	&\textbf{63.34}	 & 64.95	&\textbf{66.30	}&67.67 & \textbf{69.82}&\textbf{72.40}&71.67\\
        k & WordNetSyn& 45.31&	$\textbf{60.58}^\dag$ & \textbf{70.73}	&69.21	&61.62 & $\textbf{63.41}^\dag$&66.57	&\textbf{71.48}\\ \midrule
        % \midrule
        % Average & 52.98 & 57.13 & 60.73 & 61.76 &57.63 &58.69 &62.17 &62.61 \\ \midrule 
        l & \textbf{Average}&53.73	&\textbf{58.25}	 & 61.06 & \textbf{62.87}&61.60 & \textbf{62.68} & 64.69 & \textbf{65.37} \\ %\midrule
        \bottomrule
    \end{tabular}   
\end{table}

%% file: beir.tex
% \begin{table}[t]
% \caption{Domain adaptation comparison on BEIR Datasets for the dense bi-encoder.
% first neg are from SPLADE cocondenseur, and then for 2nd round we take the 1st round neg (I suggest to rm the 2nd round if we go for this option)}
% \label{tab:beir_adapt}
% \centering
% \begin{tabular}{c | c|c c|c c}
% \toprule
% \textbf{Dataset} & \textbf{Training} & \multicolumn{2}{c|}{\textbf{Zero Shot}} &  \multicolumn{2}{c}{\textbf{Domain Adapted}} \\ 
% & &  nDCG@10 & R@100 & nDCG@10 & R@100  \\ \midrule
% \multirow{2}{*}{Fever} & finetuning (FT) & \multirow{2}{*}{76.98} & \multirow{2}{*}{93.54} & 55.86 & 77.86  \\
% & FT+FGSM &   &  & 72.57 & 88.04  \\ \midrule
% \multirow{2}{*}{FiQA} & finetuning (FT) & \multirow{2}{*}{29.38} & \multirow{2}{*}{58.08} & 24.72 & 53.57   \\ %26.21 & 55.11 \\
%  & FT+FGSM  &   & & 23.62 & 57.35  \\ \midrule
% \multirow{2}{*}{NFCorpus} & finetuning (FT) & \multirow{2}{*}{29.17} & \multirow{2}{*}{25.35} & 38.05 &  46.43  \\ %41.34 & 47.04 \\
% & FT+FGSM &   &  & \textbf{39.42}& \textbf{48.00} \\ \midrule %\textbf{42.20} & \textbf{49.12} \\ \midrule
% \bottomrule
% % \multirow{2}{*}{TripClick} & Fine-tuning &\multirow{2}{*}{0.161} &\multirow{2}{*}{0.519} &0.218 & 0.579& \\
% % & Adapter-tuning & & &0.219 &0.578 & \\
% \end{tabular}
% \label{tab:tablesum_results}
% \end{table}

\begin{table}[t]
\caption{Domain adaptation comparison on BEIR Datasets for the dense bi-encoder.}
\label{tab:beir_adapt}
\centering
\begin{tabular}{c | c c |c c |cc }
\toprule
\textbf{Dataset} & \multicolumn{2}{c|}{\textbf{Fever}} & \multicolumn{2}{c|}{\textbf{FiQA}} &  \multicolumn{2}{c}{\textbf{NFCorpus}} \\ 
& nDCG@10& R@100 &nDCG@10 &R@100 &nDCG@10& R@100 \\ \midrule
Zero-Shot & 76.98 & 93.54 & 29.38 &58.08 & 29.17 &25.35\\ \midrule
Finetuning & 84.46& \textbf{95.78} &\textbf{32.66} & \textbf{61.75} &38.05 &46.43 \\  % 55.86 &77.86
+FGSM & \textbf{87.10} & 95.45 & 29.75& 61.69 &\textbf{39.42} &\textbf{48.00}\\ %\midrule %72.57 &88.04 \textbf{42.20} & \textbf{49.12} \\ \midrule
\bottomrule
% \multirow{2}{*}{TripClick} & Fine-tuning &\multirow{2}{*}{0.161} &\multirow{2}{*}{0.519} &0.218 & 0.579& \\
% & Adapter-tuning & & &0.219 &0.578 & \\
\end{tabular}
\label{tab:tablesum_results}
\end{table}

% \begin{table}[t]
% \caption{with neg from the dense distil niels. and then neg from 1st round (consistent if we want to have the 2 rounds)}
% \label{tab:beir_adapt}
% \centering
% \begin{tabular}{c | c|c c|c c|c c}
% \toprule
% \textbf{Dataset} & \textbf{Training} & \multicolumn{2}{c|}{\textbf{Zero Shot}} &  \multicolumn{2}{c|}{\textbf{1st Round}} & \multicolumn{2}{c}{\textbf{2nd Round}} \\ 
% & &  nDCG@10 & R@100 & nDCG@10 & R@100 &  nDCG@10 & R@100 \\ \midrule
% \multirow{2}{*}{Fever} & finetuning & \multirow{2}{*}{76.98} & \multirow{2}{*}{93.54} &  & - & -  \\
% & FGSM &   &  &  & - & - \\ \midrule
% \multirow{2}{*}{FiQA} & finetuning & \multirow{2}{*}{29.38} & \multirow{2}{*}{58.08} & 26.214 & 55.109 \\ %26.21 & 55.11 \\
%  & FGSM  &  & & 23.615 & 57.347 & - & - \\ \midrule
% \multirow{2}{*}{NFCorpus} & finetuning & \multirow{2}{*}{29.17} & \multirow{2}{*}{25.35} & 37.451 & 46.13 \\ %41.34 & 47.04 \\
% & FGSM &   &  & 19.661 & 29.355 \\ \midrule %\textbf{42.20} & \textbf{49.12} \\ \midrule
% \bottomrule
% % \multirow{2}{*}{TripClick} & Fine-tuning &\multirow{2}{*}{0.161} &\multirow{2}{*}{0.519} &0.218 & 0.579& \\
% % & Adapter-tuning & & &0.219 &0.578 & \\
% \end{tabular}
% \label{tab:tablesum_results}
% \end{table}